\documentclass{iaus}
\usepackage{graphicx}

\title[IAUS289. Trigonometric Parallax of IRAS 05168+3634] 
{The outer rotation curve project with {\sl VERA}: \\ Trigonometric parallax of IRAS 05168+3634}

\author[N. Sakai et al.]   
{Nobuyuki Sakai$^1$, Mareki Honma,$^{1, 2}$ Hiroyuki Nakanishi,$^3$\\ Hirofumi Sakanoue,$^3$ Tomoharu Kurayama,$^4$ \\
 \and the {\sl VERA} collaboration}

\affiliation{$^1$The Graduate University for Advanced Studies (Sokendai), Mitaka, Tokyo 181-8588, Japan \\ email: {\tt nobuyuki.sakai@nao.ac.jp} \\[\affilskip]
$^2$Mizusawa VLBI Observatory, National Astronomical Observatory of Japan, Mitaka,\\ Tokyo 181-8588, Japan \\[\affilskip]
$^3$Faculty of Science, Kagoshima University, 1-21-35 Korimoto, Kagoshima,\\ Kagoshima 890-0065, Japan \\[\affilskip]
$^4$Center for Fundamental Education, Teikyo University of Science, 2525 Yatsusawa, Uenohara, Yamanashi 409-0193, Japan}

\pubyear{2012}
\volume{289}  
\pagerange{119--126}
\jname{Advancing the Physics of Cosmic Distances}
\editors{Richard de Grijs \& Giuseppe Bono, eds.}
\begin{document}

\maketitle

\begin{abstract}
 We present a measurement of the trigonometric parallax of IRAS 05168+3634 with {\sl VERA}. The parallax is 0.532 $\pm$ 0.053 mas, corresponding to a distance of 1.88$^{+0.21}_{-0.17}$ kpc.
This is significantly closer than the previous distance estimate of 6 kpc based on a kinematic distance measurement. 
This drastic change in the source distance implies the need for revised values of not only the physical parameters of IRAS 05168+3634, but it also implies a different location in the Galaxy, placing it in the Perseus arm rather than the Outer arm.
 We also measured the proper motion of the source. A combination of the distance and proper motion with the systemic velocity yields a rotation velocity $\Theta$ = 227$^{+9}_{-11}$ km s$^{-1}$ at the source position, assuming $\Theta_{\rm{0}}$ = 240 km s$^{-1}$.
Our result, combined with previous VLBI results for six sources in the Perseus arm, indicates that the sources rotate systematically more slowly than the Galactic rotation velocity at the local standard of rest.
 In fact, we derive peculiar motions in the disk averaged over the seven sources in the Perseus arm of ($U_{\rm{mean}}$, $V_{\rm{mean}}$) = (11 $\pm$ 3, $-$17 $\pm$ 3) km s$^{-1}$, 
which indicates that these seven sources are moving systematically toward the Galactic Center and lag behind the overall Galactic rotation.

\keywords{Galaxy: kinematics and dynamics, ISM: individual (IRAS 05168+3634), techniques: interferometric, astrometry}
\end{abstract}

\firstsection 
\section{Introduction}
 Rotation curves can be used to determine the mass distribution in spiral galaxies by assuming force equilibrium between gravity and the centrifugal force. 
There is plenty of evidence of flat rotation curves beyond optical disks in external spiral galaxies. 
This indicates the existence of large quantities of dark matter in the outer regions of galaxies (e.g., Sofue et al. 1999).
In contrast, the shape of the rotation curve of the Milky Way is relatively ambiguous,  although it is believed to be almost flat, i.e., similar to those of external spiral galaxies (e.g., Sofue et al. 2009). 
This uncertainty is mainly owing to difficulties in measuring the relevant distances, since we are located inside the Milky Way. 
To precisely measure the distances to and proper motions of Galactic objects based on astrometry, the {\sl Hipparcos} satellite was launched in 1989 (Perryman 1989). 
However, parallax measurements with {\sl Hipparcos} were limited to distances within 100 pc from the Sun, which compares poorly to the Milky Way's disk size of $\sim$20 kpc in radius. 
Today, well-developed interferometer techniques at radio wavelengths can be used to conduct Galactic astrometry on kiloparsec scales. 
In fact, {\sl VERA} (VLBI Exploration of Radio Astrometry) and VLBA (Very Long Baseline Array) interferometry have succeeded in determining parallaxes corresponding to distances $>$ 5 kpc (e.g., Nagayama et al. 2011b; Sanna et al. 2012). 
To construct the rotation curve of the Milky Way in the outer regions with high accuracy, we used {\sl VERA} to observe Galactic objects exhibiting H$_{2}$O maser emission in the Galactic outer region. 
In this paper, we report the results for IRAS 05168+3634.

\begin{figure}[t]
\begin{center}
 \includegraphics[width=90mm]{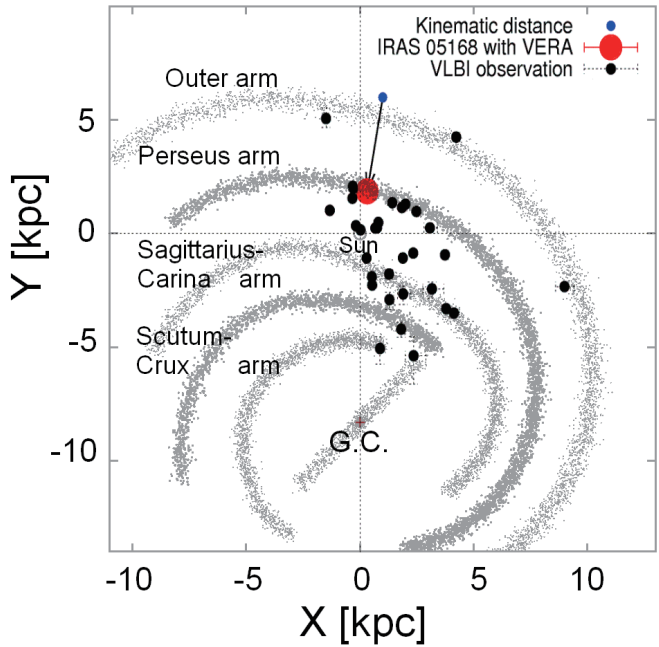} 
 \caption{Position of IRAS 05168+3634 based on both kinematic distance determination and our parallax measurement. The red circle shows our result, while the blue circle represents the kinematic distance. The black circles show previous VLBI results for other sources in star-forming regions. 
These results are superimposed on a Galactic face-on image (Georgelin $\&$ Georgelin 1976). 
The IAU value, R$_{0}$ = 8.33 kpc is assumed. The solar position is ({\sl X, Y}) = (0, 0) kpc.}
   \label{fig1}
\end{center}
\end{figure}

IRAS 05168+3634 is a high-mass star-forming region in the pre-ultracompact H{\scriptsize II} phase (Wang et al. 2009).  
An H$_{2}$O maser is detected in this region, where a CO outflow occurs (Palla et al. 1991). 
The maser position coincides with a bright red source detected with the {\sl Spitzer Space Telescope's} IRAC camera at 4.5 $\mu$m.  
Although no radio continuum (at 6cm) was detected in IRAS 05168+3634 by McCutcheon et al. (1991), Molinari et al. (1998) detected a 6cm continuum emission peak of 4.23 mJy at a position $\sim$ 1.7$'$ toward the northeast of IRAS 05168+3634. 
In this region, there are several sources at different evolutionary and star-formation phases. 
As for the distance estimate to IRAS 05168+3634, a kinematic distance of $\sim$6 kpc was obtained (Molinari et al. 1996) based on the systemic local-standard-of-rest (LSR) velocity ($v_{\rm{LSR}}$) of $-$15.5 $\pm$ 1.9 km s$^{-1}$ traced by CS (2-1) emission (Bronfman et al. 1996). 
However, this source is located at Galactic longitude $\ell$ $\sim$ 171$^{\circ}$, where kinematic distance estimates are characterized by large errors (Sofue et al. 2011a).
In fact, kinematic distances along the Sun-Galactic Center direction cannot be determined precisely, since $v_{\rm{LSR}}$ is degenerate around 0 km s$^{-1}$. 
To estimate the error in the kinematic distance at the source, we consider the error, $\Delta v_{\rm{LSR}} \simeq$ 1.9 km s$^{-1}$ for a flat rotation model. The error we obtained is large, $\sim$ 1.2 kpc or $\sim$22$\%$ at the source. Thus, VLBI observations of IRAS 05168+3634 are necessary, 
not only to precisely measure distances, but also to construct a precise outer rotation curve for the Galaxy. In this paper we present astrometric observations of IRAS 05168+3634 obtained with {\sl VERA}.

\begin{figure}[h]
\begin{center}
 \includegraphics[width=103mm]{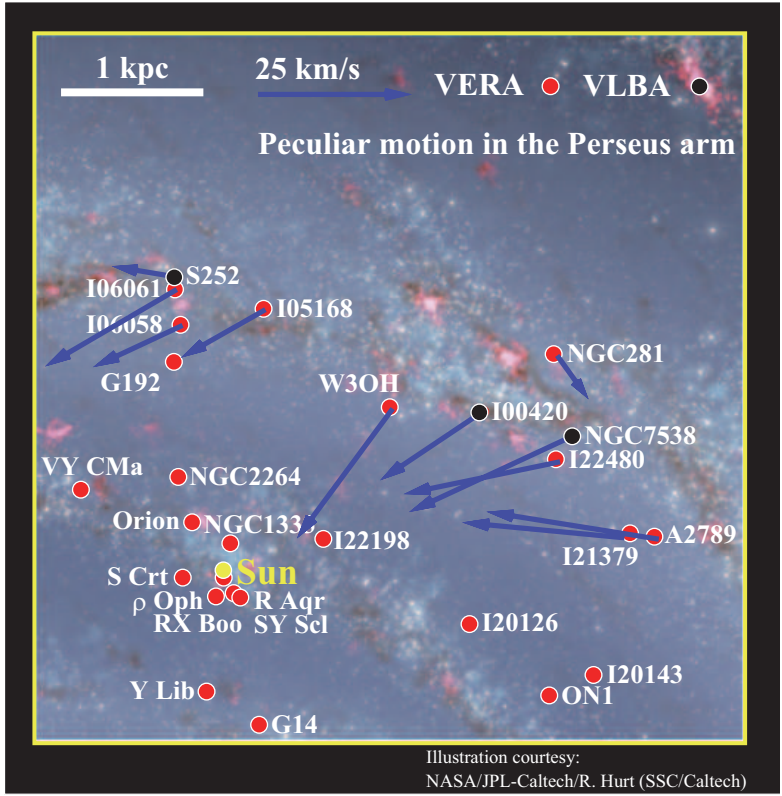} 
 \caption{Peculiar motions in the Perseus arm. The arrows represent peculiar motions for the sources located in the Perseus arm based on VLBI observations. Note that a flat Galactic rotation curve ---$\Theta$($R$) = $\Theta_{0}$--- was assumed to derive the peculiar motions. Based on the figure, almost all sources in the Perseus arm are moving systematically toward the Galactic Center and lag behind the Galactic rotation.}
   \label{fig2}
\end{center}
\end{figure}

\section{Results $\&$ Discussion}
Eleven VLBI observations with {\sl VERA} obtained between October 2009 and May 2011 yielded the trigonometric parallax and proper motion of IRAS 05168+3634.
The parallax is 0.532$\pm$0.053 mas, corresponding to a distance of 1.88$\pm$$^{0.21}_{0.17}$ kpc. The proper motion components are ($\mu_{\alpha}$cos$\delta$, $\mu_{\delta}$) = (0.23 $\pm$ 1.07, $-$3.14 $\pm$ 0.28) mas yr$^{-1}$.
The resulting distance is significantly smaller than the previous kinematic distance estimate of 6 kpc (Molinari et al. 1996). Our result places the source in the Perseus arm rather than in the Outer arm (see Fig. 1). 
Combining the distance and proper motion with the systemic velocity results in a rotation velocity of 227$^{+9}_{-11}$ km s$^{-1}$ at the source, assuming $\Theta_{\rm{0}}$ = 240 km s$^{-1}$. The result corresponding to marginally slower rotation with respect to the flat Galactic rotation curve, $\Theta$($R$) = $\Theta_{\rm{0}}$.
In addition, the slower rotation is almost consistent with previous VLBI results in the Perseus arm (see Fig. 2). \\
\ \ Fig. 2 shows the peculiar motions based on VLBI observations of the Perseus arm after subtraction of the Galactic rotation. Note that a flat Galactic rotation curve ---$\Theta$($R$) = $\Theta_{0}$--- was assumed to derive the peculiar motions. Obviously, almost all sources in the Perseus arm are moving systematically toward the Galactic Center (as shown by the positive $U$ components), and lag behind the Galactic rotation (as exemplified by the negative $V$ components). Based on the density-wave theory proposed by Lin $\&$ Shu (1964), the peculiar motions may have been generated at the inner edge of the Perseus arm, where a shock front occurs. Especially for the $V$ components in the Perseus arm, a strong correlation with galactocentric distance ($R$) has been reported (e.g., Sakai et al. 2012c). To confirm the validity of the density-wave theory, astrometry in the third Galactic quadrant for the Perseus arm is crucial since the co-rotation radius ($U$ = $V$ = 0) can exist in this region (cf. Mel'Nik et al. 1999). We have been observing Galactic star-forming regions with {\sl VERA}, which will allow us to understand not only the total mass in the Galactic disk (as part of the {\sl VERA} outer rotation curve project), but also whether the density-wave theory is correct, an assessment we anticipate to make in the near future.\\ \\

\begin{flushleft}
$\bf{Acknowledgements}$ \\
\end{flushleft}

\ NS acknowledges financial support from SOKENDAI in the form of an overseas travel grant.

\end{document}